\documentclass[aps,twocolumn,twoside,superscriptaddress,nofootinbib,floatfix]{revtex4-1}
\usepackage{amsmath,amsthm,amssymb,amsfonts}
\usepackage{standalone}
\usepackage{tikz, circuitikz}
\usepackage{graphicx}
\usepackage[binary-units]{siunitx}
\usepackage[pdftex]{hyperref}
\usepackage{xcolor}
\usepackage{pgfplotstable}
\usetikzlibrary{shapes.geometric, arrows,positioning}
\usepackage{xcolor}
\usepackage{graphicx}

\usepackage[T1]{fontenc}
\usepackage{tabularx}
\usepackage{booktabs}
\usepackage{multirow}
\newcolumntype{y}{>{\centering\arraybackslash}X}

\newcommand{\ket}[1]{| #1 \rangle}

\newcommand{\appendices}{\appendix}

\begin{document}

\title{Timing constraints due to real-time graph traversal algorithms on incomplete cluster states in photonic measurement-based quantum computing}
\author{John R. Scott}
\email{johnrscott0@gmail.com}
\affiliation{Centre for Doctoral Training in Quantum Engineering, Department of Physics, University of Bristol, UK}
\author{Krishna C. Balram}
\email{krishna.coimbatorebalram@bristol.ac.uk}
\affiliation{Quantum Engineering Technology Labs and Department of Electrical and Electronic Engineering, University of Bristol, BS8 1UB UK}

\begin{abstract}
Understanding the computational overheads imposed by classical control systems on quantum computing platforms becomes critically important as these quantum machines grow in scale and complexity. In this work, we calculate the overheads imposed by the implementation of real-time graph traversal algorithms needed to find computational paths through incomplete cluster states for the implementation of one-qubit gates; a necessary requirement for a realistic implementation of photonic measurement-based quantum computing. By implementing two different algorithms, a global breadth-first search that searches the entire cluster state and an incremental version that traverses a narrow sub-section of the cluster state, we analyze the tradeoff between the accuracy of finding viable paths and the speed at which this operation can be performed, which constrains the overall photonic clock cycle of the system. We also outline the broader implications of our results for implementing classical control systems for measurement-based photonic quantum computing.  
\end{abstract}

\maketitle

\section{Introduction}
\label{sec:introduction}

As quantum computing platforms grow in scale and complexity, it has become increasingly clear that the classical control infrastructure required to support these quantum machines must keep up in performance and sophistication. This is best illustrated by Google's ground-breaking quantum supremacy experiment~\cite{Arute2019}, where the final experimental run took $\approx$ \SI{200}{\second}, but the classical calibration and control needed to get the quantum processor ready for this data collection run required $\approx\SI{36}{\hour}$ for the first cool-down, and \SI{4}{\hour} per day thereafter.

A more interesting classical overhead is caused by classical computations that need to be performed in real time, while the quantum processor is running. This is required, for example, in implementations of measurement-based photonic quantum computing (MBQC)~\cite{briegel2009measurement} or error-correction protocols~\cite{krinner2022realizing,fowler2012surface}. In both cases, measurement outcomes are processed and used to perform subsequent operations on the quantum state in order to ensure that state evolves deterministically. These classical calculations must be precisely quantified and their run times characterised through analysis of concrete control system architectures, because they may severely constrain the operation of the quantum processor.

\begin{figure*}[t]
  \centering
  \includegraphics[width=0.9\textwidth]{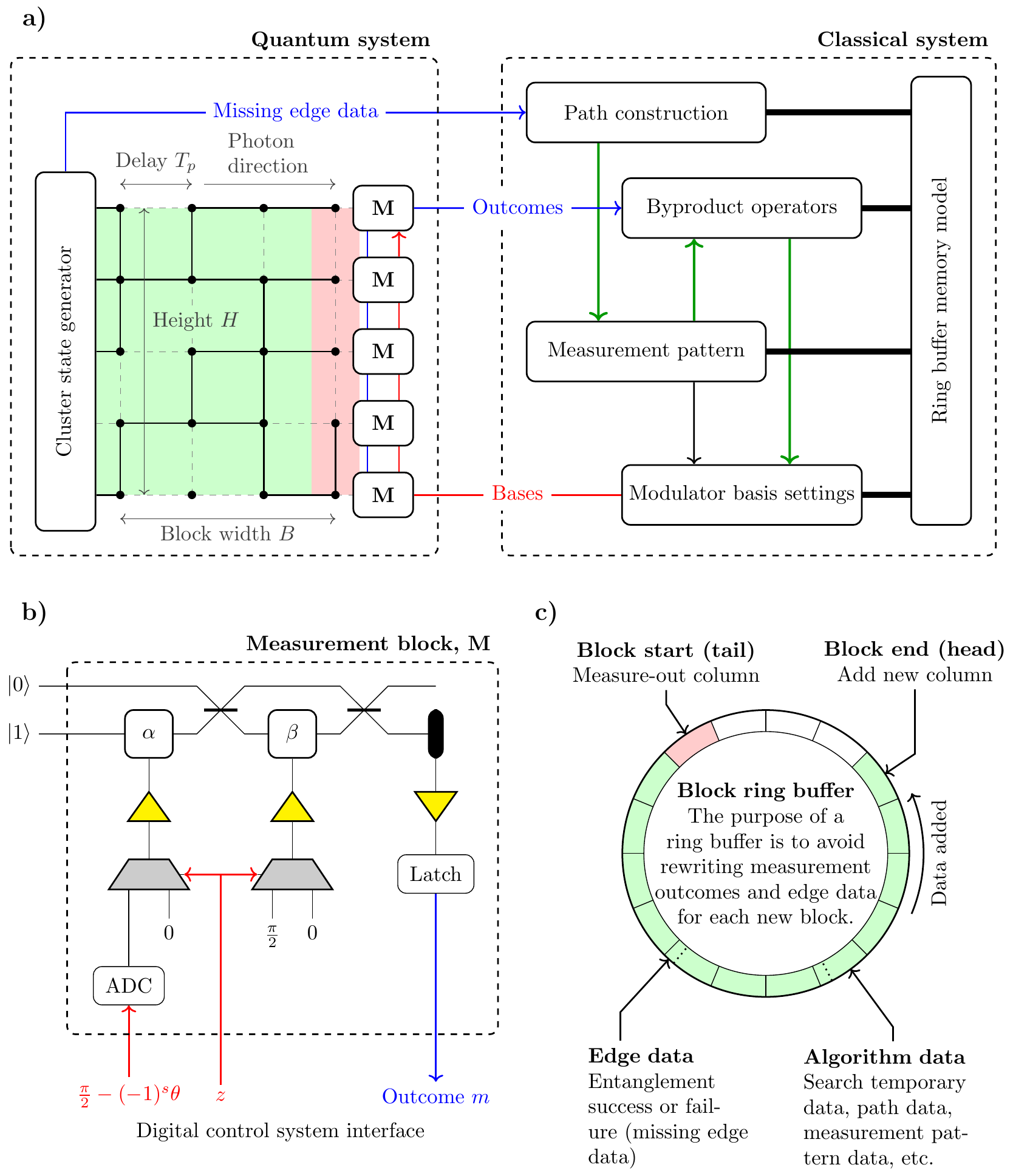}  
  \caption{a) Schematic showing the interface between the photonic quantum computer and the classical control system. The quantum computation proceeds by single qubit measurements (M) of the incomplete cluster state on the left, where successful entanglement is represented by solid block lines. The classical control system needs to calculate a path through this incomplete cluster state to generate the desired measurement pattern (which implements a one-qubit gate) and update the byproduct operators based on the measurement outcomes. These classical computations need to be performed before setting the bases for the next round of measurements and our main focus in this work is to estimate the timing overheads these calculations place on the photonic clock cycle. The arrows in the classical system show the computational dependencies, with the green arrows highlighting the critical timing path. (b) Schematic of the measurement block which implements the (discrete-variable, dual-rail-encoded) photonic-qubit basis measurement. The value of the $z$ bit selects the angle in the multiplexer (grey trapezium), which is then amplified (yellow triangle) to set the bias voltage on the modulator. The basis angle is generated using an analog-to-digital converter (ADC) (c) The ring buffer data structure used to store all the graph data (entanglement edges), and the node data generated by the graph traversal algorithms. A detailed discussion can be found in Appendix~\ref{sec:ring-buffer-model}.}
  \label{fig:full-system}
\end{figure*}

Measurement-based photonic quantum computers provide an ideal test platform to study these overheads, since the upper bound on the lifetime of a qubit in integrated platforms is defined by the length of an on-chip delay line, on the order of a few nanoseconds~\cite{zhou2018integrated}. As we showed in our previous work~\cite{scott2022timing}, even a relatively idealized photonic quantum computer with a defect-free cluster state shows significant timing constraints when the control system is implemented with a state-of-the-art field-programmable gate array (FPGA), because of the need to track classical bits (byproduct operators) and implement measurement-induced feedforward operations~\cite{Raussendorf2003}. In that work, most of the timing constraints were of hardware origin, with the latency originating primarily from the finite speed of the digital logic inside the FPGA, rather than the complexity of the algorithms being implemented. In this work, we extend our analysis to the case of imperfect cluster states, as would be realistically encountered in any physical implementation of MBQC, and ask what additional constraints arise when the classical control needs to perform more sophisticated real-time calculations~\cite{Kieling2007,morley2017physical,herr2018local}.

Our main aim in performing this detailed analysis is to understand the scale of the classical overhead introduced by the need to perform path-finding algorithms in real-time, i.e., within one photonic clock cycle. This analysis can then be used to help determine attributes of the control system hardware, such as the memory architecture~\cite{mittal2021survey}, and the design of the control path for the system~\cite{hennessy2011computer}. An analysis like this is the only way to understand whether off-the-shelf electronics (like fast FPGAs) can meet the requirements of photonic MBQC, or if custom application specific integrated circuits (ASICs) are the only way forward. 
\begin{figure*}[t]
  \centering
  \includegraphics[width=0.8\textwidth]{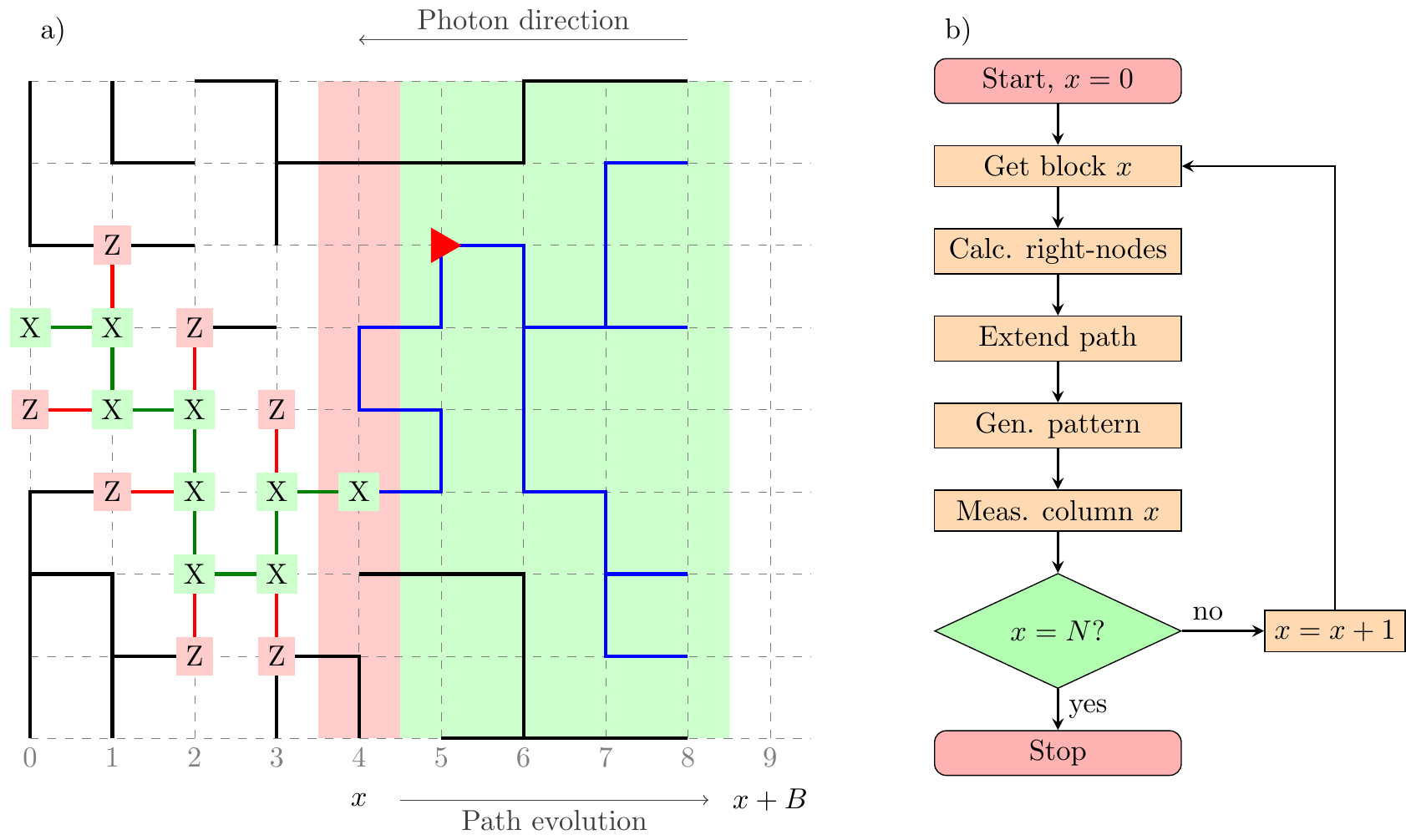}
  \caption[Block search implementation of photonic MBQC]{(a) Diagram of the incomplete cluster state being searched. Compared to Figure~\ref{fig:full-system}a, the direction of photon movement is reversed so that the path can be depicted as evolving from left to right. The one-qubit path is shown by the green line, and red lines show entanglement links that must be removed (cut out). The diagram shows the one-qubit identity operation, where all on-path qubits are $X$ measurements. (b) Diagram of the algorithm for finding path extensions and generating the measurement pattern. The incomplete cluster state is searched one block at a time (the block starting at column $x$) to establish a path for the qubit. The search process is decoupled from the path extension through the establishment of right nodes on the path (see Section~\ref{sec:search-viable-paths}). Once the path is established, the measurement pattern is generated, taking account of qubits which must be cut out around the path; then the left-most column of the block is measured out. In the implementation considered here, data relating to the nodes in the shaded green region is stored in a ring buffer, as described in Appendix~\ref{sec:ring-buffer-model}.}
  \label{fig:block-search}
\end{figure*}

We use software emulation of the system shown in Figure~\ref{fig:full-system} (described in Section~\ref{sec:mbqc}) to track key memory-related metrics that can be directly translated into timing constraints (in Section~\ref{sec:timing-impl-phot}), relevant to the operational speed of a photonic quantum computer. A thorough description of the classical algorithms we emulate is given in Sections~\ref{sec:global-breadth-first} and \ref{sec:incr-breadth-first}. This analysis forms a prerequisite step before realising a full digital design, which would afford a complete analysis of all classically-imposed timing constraints. Our analysis methodology (emulating instances of the classical algorithms and counting memory operations) can be readily extended to any photonic quantum computing architecture, or other settings in which real-time closed-loop control of quantum states is required.

\section{The system model}
\label{sec:mbqc}

\begin{figure*}[t]
  \centering
  \includegraphics[width=0.85\textwidth]{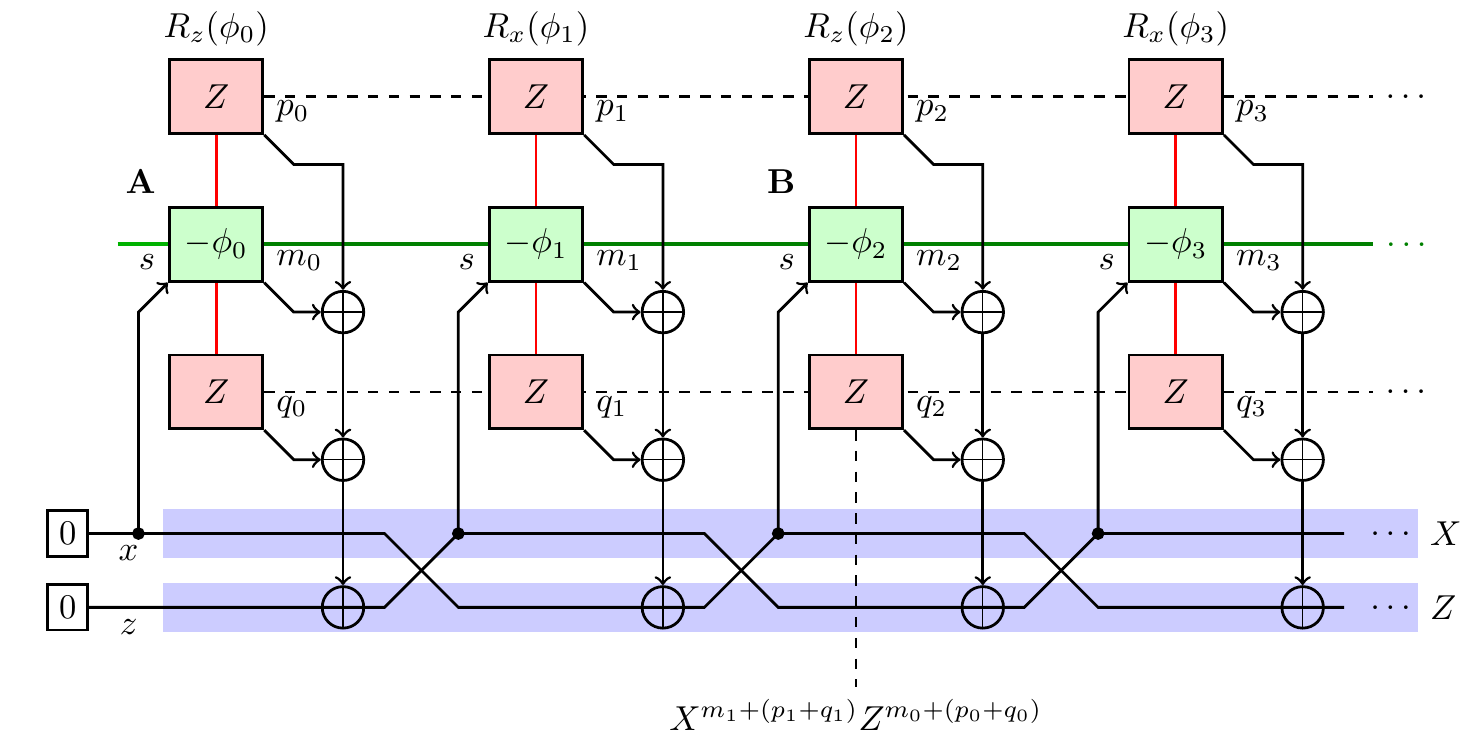}  
  \caption[One-qubit-gate measurement pattern including cut-outs]{Schematic implementation of an arbitrary one-qubit gate mapped on to a 2D cluster state, showing the sequence of operations. The green line encodes the logical one-qubit gate, in this case along a horizontal line. The adjacent rows of the cluster state are removed (cut out) using $Z$ measurements (shown in red). The byproduct operators swap every other column, as shown in the blue trace at the bottom, for purposes relating to the quantum simulation of the measurement pattern (see Appendix~\ref{sec:simul-volt-noise}). The circles denote (classical) bitwise XOR operations. The computations shown here are also representative of mapping the one-qubit gate onto an arbitrary path through an arbitrarily connected cluster state. The byproduct operators display the same behaviour, alternating with incrementally increasing path index $n$. Measurement outcomes for any cut-out cluster qubits adjacent to a path qubit $q_n$ are added to the measurement outcome from $q_n$, before this is added to the byproduct operator $z_{2k}$. How this pattern is mapped onto an arbitrary path is shown in the algorithm in Figure~\ref{fig:pattern-gen}.}
  \label{fig:cut-out-pattern}
\end{figure*}

This paper concerns the implementation of one-qubit gates along paths through incomplete cluster states in a model of MBQC. Incomplete cluster states arise in photonic quantum computing due to the probabilistic nature of entangling gates~\cite{kok2010introduction}. The resulting incomplete cluster state, with missing edges, is the underlying quantum resource for photonic MBQC~\cite{Kieling2007}. Figure~\ref{fig:full-system} shows the model of quantum computing system we consider in this paper.

The interface between the classical control system and the quantum system is shown in Figure~\ref{fig:full-system}a. A cluster state generator creates an incomplete 2D cluster state of height $H$ and block width $B$, which is the number of columns of photons ``alive'' at any given time (the shaded green/red block).

The classical control system (Figure~\ref{fig:full-system}a, right) receives graph edge data from the cluster state generator, and measurement outcomes from the measurement blocks $M$, and uses this information to implement MBQC measurement patterns~\cite{Raussendorf2003,quant-ph/0603226}. This paper is devoted to the implementation of one-qubit gates mapped onto paths through this incomplete cluster state. To perform this task, in each photonic clock cycle, the control system must:
\begin{enumerate}
\item Find a way to extend the logical one-qubit path into the newly generated column of the cluster state.
\item Map a measurement pattern to the path, containing the rules for how to implement a one-qubit gate.
\item Compute measurement basis settings required for the next measurement round.
\item Use measurement outcomes to update the byproduct operators.
\end{enumerate}
Of these steps, (1) and (2) are performed after a new column of photons is added by the cluster state generator, step (3) is used to obtain measurement bases to measure out a column of photons on the right (the red column), and (4) involves processing the measurement outcomes into byproduct operators and generating information about how to set the next round of measurement bases.

All these steps must occur within the photonic clock cycle, $T_p$. While the computation is taking place, the photons are stored in a $T_d = BT_p$ delay line (either on-chip or in optical fibre), of length proportional to the block width $B$. This width must be large enough to ensure that there is a reasonable chance of finding paths, by providing a sufficient buffer region for the control system algorithms to operate correctly.

The memory model for the control system, which is the basis for the timing results in this paper, is shown in Figure~\ref{fig:full-system}c. It is based on a ring buffer, an implementation of a first-in, first-out data structure, which allows new (column) data to be added without having to move all the data already in the buffer~\cite{kruse1998data}. Therefore, the insertion time into the structure when a new column of photons is generated is $O(H)$, an essential requirement for an efficient control system. Each segment shown in Figure~\ref{fig:full-system}c includes all the data for a particular column, comprising the edge data, and algorithm data which may include flags, distances, offsets, and other algorithmic variables.

Each measurement is performed by a measurement block, shown in Figure~\ref{fig:full-system}b, which measures a (discrete-variable) photonic qubit in the dual-rail encoding. Two modulator angles are necessary: $\alpha$, which determines the $xy$-plane angle of the measurement; and $\beta$, which chooses between a computational basis measurement and an $xy$-plane measurement. The system involves high-speed analog devices (amplifiers, latches, analog-to-digital converters, and latches), which we do not consider in this paper, but which must operate fast enough to meet $T_p$ minus the time taken up by the classical computation steps (the classical overhead).

\subsection{The one-qubit measurement pattern}
\label{sec:dynam-meas-patt}

\begin{figure}[t]
  \centering
  \includegraphics[width=0.9\columnwidth]{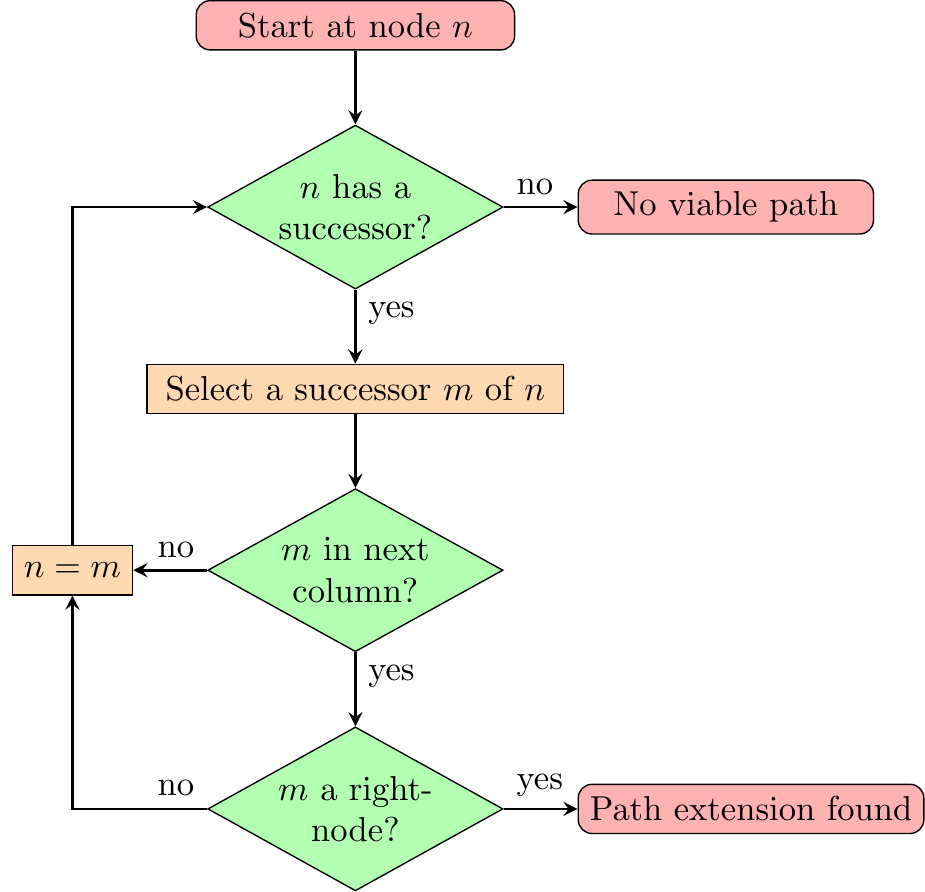}
  \caption[Algorithm for extending the path]{Algorithm for extending the path to a right node in the next column. The algorithm advances along a potential path (the blue lines in Figure~\ref{fig:block-search}a), and picks a successor at random whenever a branch point is encountered. The algorithm stops when a right node is encountered, which guarantees that the leftmost column of the green block can be safely removed without compromising the path.}
  \label{fig:path-extend}
\end{figure}

\begin{figure}[t]
  \centering
  \includegraphics[width=0.8\columnwidth]{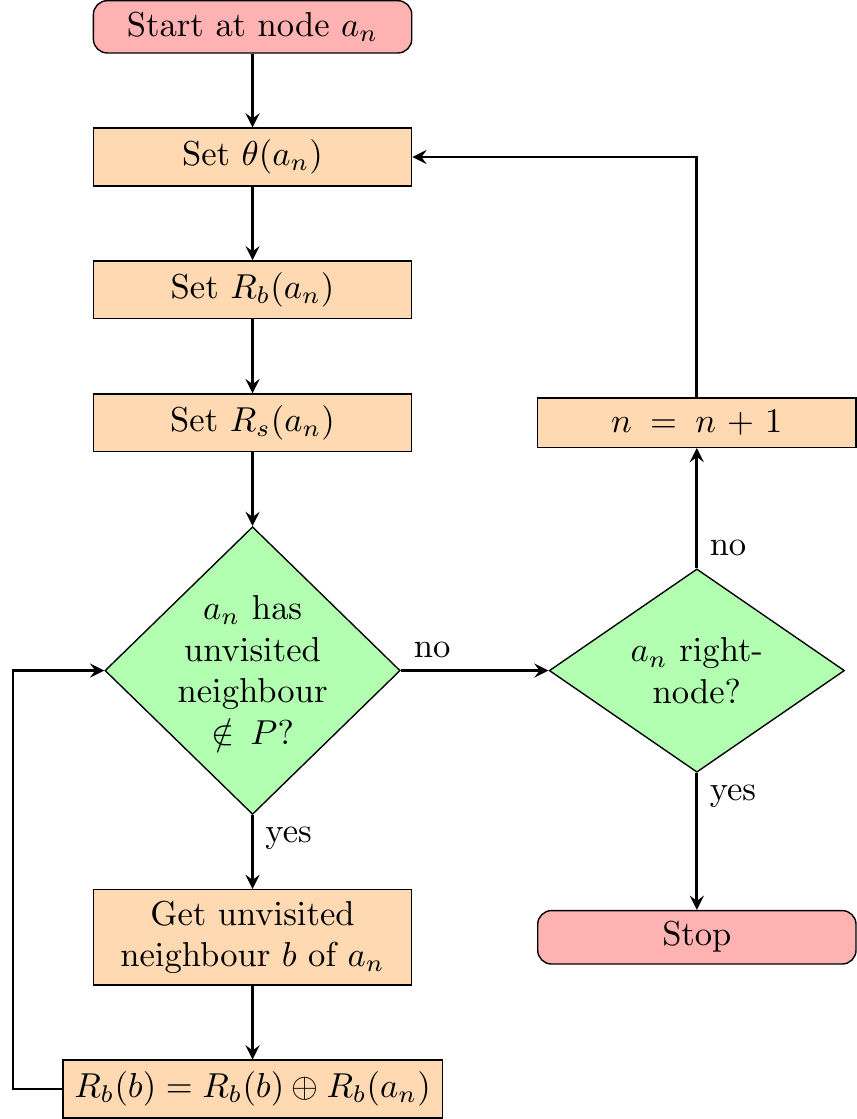}  
  \caption[Algorithm for generating local pattern rules]{Algorithm for generating local pattern rules on a path extension (up to the next right node). The path extension is traversed once, forwards. Here, $\theta$ is the base angle, $R_{b}$ is the byproduct operator update rule, and $R_s$ is the rule for computing the adaptive measurement setting. The set of on-path nodes is abbreviated $P$. The presence of neighbouring qubits around the on-path node $a_n$ leads to an update in $R_b$, to account for the effect of cut-out qubits (the ``unvisited'' status of the cut-out qubits only applies to the innermost loop; any given cut-out qubit may be visited more than once from different on-path qubits). As an example, for the identity pattern, all on-path measurements are $X$: $\theta(a_n)=0$, $R_s(a_n)=(0,0)$, and $R_b(a_n) = (n+1\text{ mod }2,n\text{ mod }2)$. Further information about the local pattern rules is provided in Appendix~\ref{sec:local-pattern-rules}.}
  \label{fig:pattern-gen}
\end{figure}

The measurement pattern implemented by the system is the arbitrary one-qubit gate along a path of edges through the cluster state. In the special case that the path is horizontal and the cluster state is fully connected, this pattern is shown in Figure~\ref{fig:cut-out-pattern}. Red boxes containing $Z$ are the computational basis measurements needed to disconnect nearby entangled qubits from the path (cut out qubits). Green boxes containing an angle $\theta$ are measurements in the $xy$-plane of the Bloch sphere at an angle $(-1)^s\theta$ to the $x$-axis\footnote{This measurement can be implemented in the dual-rail encoding by setting $\alpha=\pi/2-(-1)^s\theta$ and $\beta=\pi/2$ in Figure~\ref{fig:full-system}b.}. The choice of this angle determines what logical one-qubit operation is performed. The individual measurement outcomes $m_n,p_n,q_n$ from cluster qubits (either 0 or 1) are used to calculate the byproduct operators, which are then used for determining the adaptive measurement settings $s$. Provided that the pattern terminates on an even-indexed column $n=2k$ (counting from zero), and the $\theta$ angles are chosen as shown in Figure~\ref{fig:cut-out-pattern} (note the minus signs), the following arbitrary one-qubit gate is implemented:
\begin{equation}
  \label{eq:one-qubit-pattern-u}
U_{2k} = \big(R_x(\phi_{2k-1})R_z(\phi_{2k-2})\big)\dots \big(R_x(\phi_{1})R_z(\phi_{0})\big).
\end{equation}
The interpretation of this pattern (for the example $n=2$) is as follows, assuming that the cluster state is $3\times 3$: if the first qubit (marked \textbf{A} on Figure~\ref{fig:cut-out-pattern}) began in the state $\ket{\phi_\text{in}}$, before it was entangled with the rest of the cluster state, then the final qubit \textbf{B} (the only qubit that is left unmeasured) would be in the state
\begin{align}
  \ket{\psi_\text{out}} &= X^xZ^zU_2\ket{\psi_\text{in}}\nonumber\\
  &=X^{m_1+(p_1+q_1)}Z^{m_0+(p_0+q_0)}R_x(\phi_{1})R_z(\phi_{0})\ket{\psi_\text{in}}\label{eq:one-qubit-example}.
\end{align}
The byproduct operator term $X^xZ^z$, which may be read out beneath the terminating column $n=2$ as shown in Figure~\ref{fig:cut-out-pattern}, is an intrinsic feature of measurement patterns, and may be corrected after measuring the final state from the quantum computation~\cite{Raussendorf2003}. However, adaptive measurement settings are derived from byproduct operators, so they must be calculated in real time while the measurement pattern is being executed.

It is important to keep in mind that the logical one-qubit gate $X^xZ^zU_2$ is applied to the logical state $\ket{\psi_\text{in}}$ to produce the output state $\ket{\psi_\text{out}}$, which evolves as the cluster qubit measurements are performed, but is not associated with any particular cluster qubit in the pattern. The logical $R_x$ and $R_z$ rotations implemented in Equation~\eqref{eq:one-qubit-example} are distinct from the one-qubit operations applied to cluster qubits to set the $xy$-plane or $Z$ measurements.

Rigorous information about how to derive measurement patterns is available in~\cite{Raussendorf2003}; a quicker tutorial introduction is~\cite{quant-ph/0603226}. Measurement patterns are more easily derived in practice using the ZX-calculus, as described for example in~\cite[Chapter 6]{van2020zx}.

This pattern must be mapped onto a path through the cluster state. The process of finding this path is discussed in the next section.

\subsection{Path search}
\label{sec:search-viable-paths}

\begin{table*}[t]
  \centering
  \begin{tabularx}{0.85\textwidth}{l l} 
    \toprule
    Ring-buffer data & Meaning\\
    \midrule
    distance, $d(n)$ & The distance from the root node to node $n$ \\
    predecessor, $\text{pred}(n)$ & The unique BFS predecessor of $n$ \\
    inaccessible flag (IBFS only) & A Boolean flag to indicate whether $n$ is an inaccessible exit node \\
    right-node flag & A Boolean flag to indicate whether $n$ is a right node \\
    successors, $\text{succ}(n)$ & The set of successors of $n$ \\
    \midrule
    Other notation and terms & \\
    \midrule
    $n.x$ and $n.y$ & The $x$ and $y$ coordinates of the node $n$ in the cluster state\\
    $\text{pop}(Q)$ & Get the next element from the queue $Q$ \\
    root node & The starting point for the current path extension (in the red column)\\
    \texttt{F,A} & Two additional flags required for the implementation of the algorithms\\
    $x,y$ & Two integer variables used for the implementation of the algorithms\\
    \bottomrule
  \end{tabularx}
  \caption[Data required for GBFS and IBFS]{Local data required for the implementation of GBFS and IBFS, and notation used in Figures~\ref{fig:gbfs} and \ref{fig:ibfs}. In traversing the graph forwards from the starting node, the distance to each node is recorded. In addition, each node apart from the starting node stores a predecessor, the node from which it was visited. These predecessors are reversed to generate a successor set for each node, which is used in the path extension algorithm (see Figure~\ref{fig:path-extend}). Finally, a flag is used to indicate whether a given node is a right node. For IBFS, an additional flag is required as part of the failed-path pruning step.}
  \label{tab:gbfs-data}
\end{table*}

After a new column of photons has been generated, an algorithm is required to quickly extend the one-qubit path and pattern from column $x$ to column $x+1$ of the green block in Figure~\ref{fig:block-search}a, so that the next red column may be measured and removed. Following~\cite{morley2017physical}, we divide the pathfinding process into a \textit{search process}, which establishes \textit{potential paths} extending all the way to the rightmost (new) column of the green block, and a \textit{path extension} phase, which selects one path out of the set of available paths (shown in blue in Figure~\ref{fig:block-search}a).

The simpler process is the path extension, which is shown in Figure~\ref{fig:path-extend}. It consists of stepping along potential paths, making random choices at each branch point, until it reaches a \textit{right node} (shown as a red filled triangle in Figure~\ref{fig:block-search}a); this is the first node along the path having the property that all path successors lie strictly\footnote{Allowing path successors above and below is excluded to enforce the uniqueness of the right node in each column.} to the right of it. If the path is not advanced to a right node, then the measurement of the leftmost column of the green block will potentially cut off the path so that it is not possible to extend it further to the right, thereby ending the quantum computation. The calculation of right nodes, which must be performed before extending the path, intrinsically involves a traversal of the graph backwards, as we describe in Sections~\ref{sec:global-breadth-first} and~\ref{sec:incr-breadth-first}.

\subsection{Pattern generation}
\label{sec:pattern-generation}

Once the search process is complete, the measurement pattern must be mapped to the path extension, before measurement settings can be derived. An algorithm suitable for this purpose is shown in Figure~\ref{fig:pattern-gen}. The algorithm outputs local rules, specifically designed to be compatible with the ring-buffer structure in Figure~\ref{fig:full-system}c, which control how measurements are performed when the red column is measured out.

The meaning of ``local'' is that each node in the column to be measured contains rules that specify in what basis to measure it, and what to do with the measurement outcome, without needing to refer to rules or outcomes from surrounding nodes. This maximises the efficiency of the measurement process, reducing data dependencies and making it parallelisable. In addition, per-node data is easily stored in the ring buffer structure.

The rules can be briefly summarised as follows. Each node $a_n$ along the path (indexed by $n$) contains a value $\theta(a_n)$ which sets the $xy$-plane measurement basis of this node. A rule $R_s(a_n)$ describes how to compute the adaptive measurement setting for $a_n$ from the current byproduct operators. Finally, $R_b(a_n)$ describes how to update the byproduct operators from the measurement outcome from $a_n$. Cut-out qubits surrounding the path also have rules specifying how to compute byproduct operators in line with Figure~\ref{fig:cut-out-pattern}. More information about how the rules are defined is contained in Appendix~\ref{sec:local-pattern-rules}.

The cost of this algorithm must also be accounted for in the digital processing time, although we have not performed this analysis in this paper because it is substantially less than the main bottleneck involved in the classical computation. This is the search process, for establishing potential paths and right nodes, which is discussed in the next section.

We verified that the pattern rules do indeed enable the encoding of a one-qubit measurement pattern, by performing a quantum simulation of the entire scheme. This is described in Appendix~\ref{sec:simul-volt-noise}.

\section{The two search algorithms}
\label{sec:two-search-algor}

\begin{figure*}[t]
  \centering
  \includegraphics[width=0.7\textwidth]{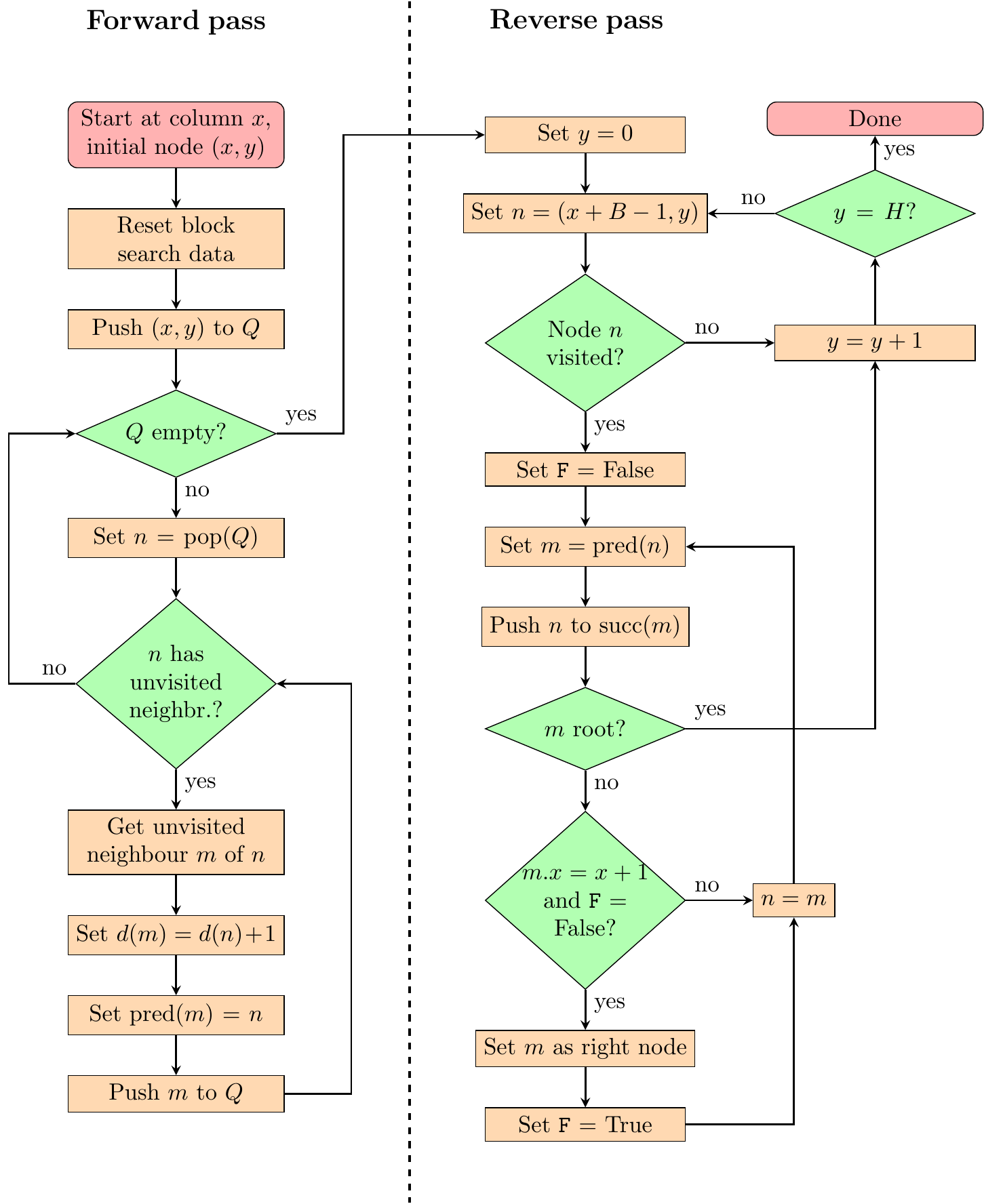}
  \caption[Global breadth-first search (GBFS) algorithm]{Flowchart outlining the GBFS algorithm, which searches block $x$ (beginning at column $x$; see Figure~\ref{fig:block-search}a). The forward pass is a standard BFS used to construct a predecessor tree. The reverse pass traverses backwards through this predecessor tree to establish right nodes for path extension.}
  \label{fig:gbfs}
\end{figure*}

The main timing analysis derived in Section~\ref{sec:timing-impl-phot} is based on the implementation of two search algorithms, for finding potential paths through the cluster states. The first, the global breadth-first search (GBFS), is a na{\"i}ve algorithm based on repeated breadth-first search of the active area (green and red block in Figure~\ref{fig:block-search}a). This algorithm works all the time (provided the edge probability in the cluster state is high enough), but is very inefficient. The second algorithm, the incremental breadth-first search (IBFS) attempts to solve the efficiency problem of the GBFS algorithm by using search information from previous searches to cut down the run time of the algorithm. In Section~\ref{sec:emul-search-algor}, where we emulate the algorithms, we show that IBFS fails in nearly all cases. However, due to its speed, it forms a more promising basis for investigation into better search algorithms.

The details of the GBFS and IBFS algorithm are outlined in the following sections.

\subsection{Global breadth-first search}
\label{sec:global-breadth-first}

The GBFS algorithm is shown in Figure~\ref{fig:gbfs}. The algorithm begins after a new column of photons has been generated, and edge data has been recorded in the ring buffer. It comprises a forward breadth-first search over the nodes in the ring buffer, which calculates distance and predecessor information, followed by a reverse pass which calculates the successors (that form candidates for the path extensions) and the right nodes. This information is stored at each node in the ring buffer. The data used by GBFS (and IBFS; see Section~\ref{sec:incr-breadth-first}) is summarised in Table~\ref{tab:gbfs-data}.

\subsubsection{Breadth-first search}
\label{sec:breadth-first-search}

The first step of the algorithm is a breadth-first search (BFS). The BFS begins at a particular node on the path $(x,y)$, in the left-most of the block (the red column labelled $x$ in Figure~\ref{fig:block-search}a). First, the algorithm must reset all the local data (see Table~\ref{tab:gbfs-data}) in the ring buffer, which constitutes temporary information from one photonic cycle to the next.

Next, a standard implementation of the BFS algorithm~\cite{cormen2022introduction} is used to construct a tree of predecessors, and each node is assigned a distance $d$ that is one greater than its predecessor. A queue (first-in first-out) structure $Q$ is used to maintain the breadth-first order of traversal of the nodes.

\subsubsection{Reverse pass and right node calculation}
\label{sec:reverse-pass-right}

The tree of predecessors is used in the reverse pass of the algorithm to iterate from exit nodes (visited nodes in the right-most column of the block) back to the root node. At each node, the predecessor relationship is recast as a successor relationship, which forms the basis for path extensions. It is important to note that this step cannot be optimised away -- it is not possible to obtain path extensions using local predecessor information, because there is no (local) way to obtain viable successors from a given node, based only on the predecessors information. The right node along each potential path is identified as the first node encountered in column $x+1$ during the reverse pass (see the red triangle in Figure~\ref{fig:block-search}a, found by traversing the blue tree backwards).

\subsubsection{Performance problems in GBFS}
\label{sec:perf-probl-gbfs}

Local data for all the nodes must be unconditionally cleared at the beginning of the GBFS algorithm, resulting in a lower bound of $HB$ writes to those memory locations. Then, with high probability (depending on the edge probability $p$), a high proportion of the block nodes are visited again and assigned predecessor and successor information, much of which likely duplicates the data that was already there before it was cleared. All these writes have to happen in the timescale of a single photonic clock cycle.

This is the primary motivation for developing an alternative, such as the IBFS algorithm in the next section. However, attempting to reuse the data in the block is not as simple as it may appear, We discuss the major failure of the IBFS algorithm, and how it relates to the resetting of the data in the ring buffer, in Section~\ref{sec:other-failures}.

\subsection{Incremental breadth-first search}
\label{sec:incr-breadth-first}

\begin{figure*}[t]
  \centering
  \includegraphics[width=0.9\textwidth]{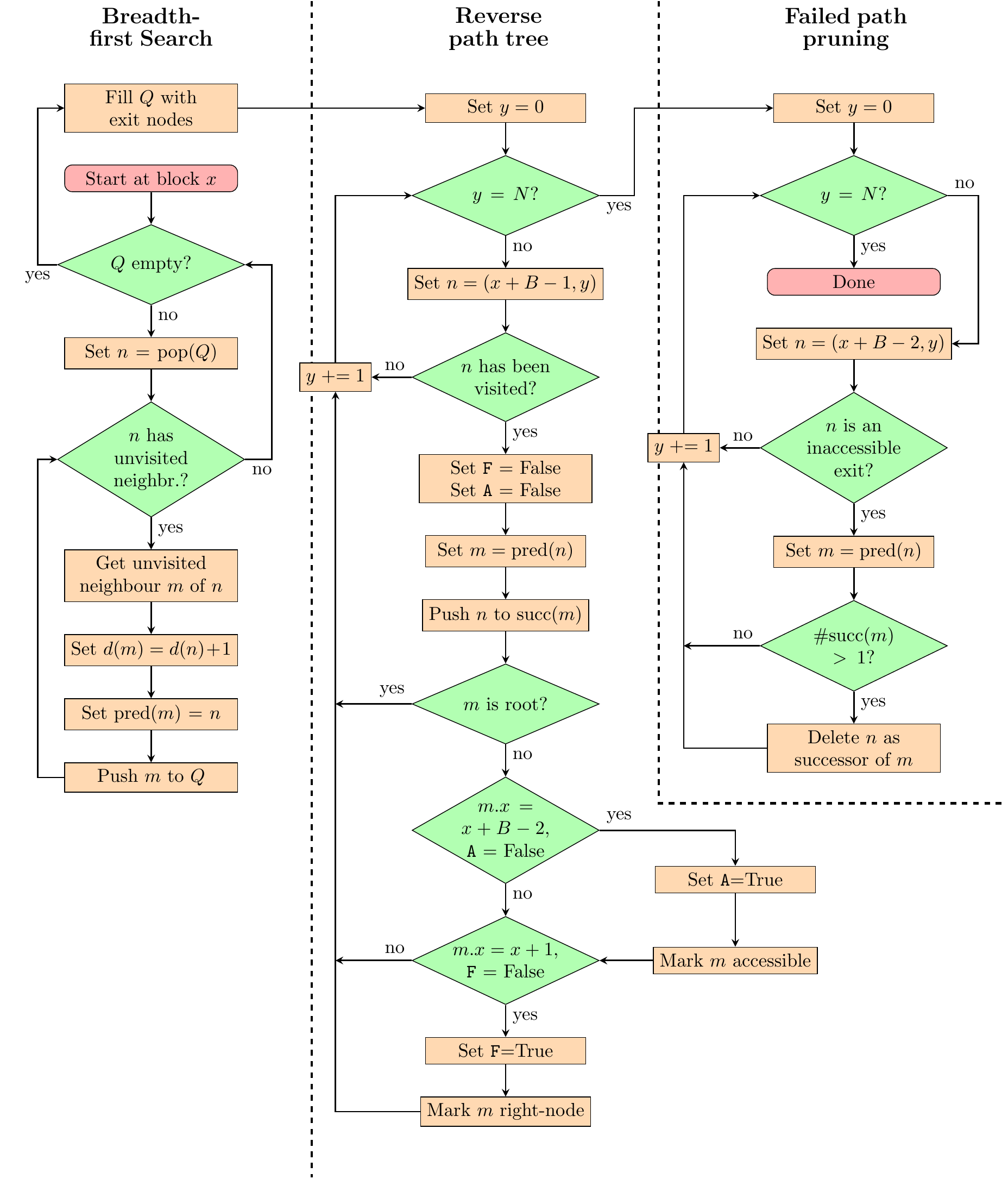}  
  \caption[Incremental breadth-first search (IBFS) algorithm]{Flowchart outlining the IBFS algorithm. The IBFS algorithm shares a lot of similarity with the GBFS algorithm, with the key differences being the reduced search volume, the reuse of predecessor tree data, and the associated need to perform failed path pruning (right).}
  \label{fig:ibfs}
\end{figure*}

The IBFS algorithm is the simplest possible attempt to remove the main defect of GBFS -- the resetting of all the search data at the beginning of each new clock cycle. The algorithm begins after a new column of photons has been generated, and edge data has been recorded. However, this time, only the region between the penultimate column and the right-most column of the block is searched. This significantly reduces the amount of the graph that must be traversed each photonic clock cycle. However, as a result of keeping the data from previous searches, several changes must be made in the reverse pass of the algorithm.

The IBFS algorithm is shown in Figure~\ref{fig:ibfs}, and described in the sections below. Local data used in IBFS is shown in Table~\ref{tab:gbfs-data}. This includes a new flag, to mark when an exit node becomes inaccessible. This relates to the main new feature of IBFS compared to GBFS -- the need to prune failed paths.

The implementation of BFS in this algorithm is quite similar to the version in GBFS. However, it is not necessary to reset the data in the ring buffer, because the main purpose of IBFS is to reuse the contents of the buffer.

Secondly, the BFS does not start with just one root node; instead, it begins with all the exit nodes from the previous block. These are the visited nodes that were in the right-most column of the previous block, and are now in the penultimate column due to the newly added column of photons. These nodes are already assumed to be in the queue from the previous iteration of the algorithm\footnote{For simplicity, we do not consider in detail how all the edge-cases for these algorithms are implemented (for example, the initial block and final block of the window). The interested reader should consult the code for MBQCSIM, which is publicly available.}. This is the main ``incremental'' feature of the algorithm; only the new columns on the right are searched each photonic clock cycle.

As a result of the need to begin each search with the queue populated by exit nodes, it is necessary to fill $Q$ with the exit nodes at the end of the search process. This is achieved by looping over the right-most column of the block and pushing any visited node to the queue.

Although it would appear that IBFS performs an identical search process to GBFS, albeit over several photonic clock cycles instead of one, the two algorithms are not equivalent. Not only may they produce different predecessor relationships; it is not even necessarily the case that they will assign the same distances to nodes. This is because a newly added column on the right may expose a shorter path to an \textit{already visited} node inside the block. Because nodes are only ever visited once (in GBFS they may be visited once \textit{per clock cycle}), the distances are not rewritten. This is not a problem, because shortest paths are not an important criterion in our analysis of photonic MBQC.

The main problem with IBFS is the possibility that a path may be invalidated when a new column is added -- for example, if it turns out that path leads to a dead-end. This problem is addressed in Section~\ref{sec:failed-path-pruning}.

\subsubsection{Reverse pass and right node calculation}
\label{sec:reverse-pass-right-ibfs}

Like the GBFS algorithm, it is necessary to perform a reverse pass over the block in order to establish right nodes. Due to the incremental nature of the algorithm, it is only necessary to traverse the predecessor paths up to an exit node in column $x+B-2$ (the penultimate column), on the grounds that a previous iteration of IBFS will have established successor information before that point.

However, it is not possible to compute right nodes in this way. Although one could try to establish right nodes in column $x+B-2$, by marking the first node in column $x+B-2$ a right node, this will not work, because it is highly likely the path may backtrack into the left region of the block via a path not yet visible to the algorithm (because those photonic columns have not been created yet). As a general rule, we found it is best to calculate right nodes at the left side of the block, because this maximises the forward path length on which the right node is based. Therefore, it is still necessary to make at least one reverse pass over the entire block, even though no full forward pass is necessary. This is still simpler than GBFS, because the reverse pass only involves checking for right nodes, not writing all the successor information (as we show in Section~\ref{sec:emul-search-algor}).

\subsubsection{Failed-path pruning}
\label{sec:failed-path-pruning}

The most important new part of the algorithm is the need to prune failed paths. Failed paths arise because a string of successors established during the searching of block $x$ may become invalid when block $x+1$ is searched, if the path leads to a dead-end. This cannot happen in the GBFS algorithm, because the ring-buffer is reset at the start of each block search.

To establish failed paths, it is necessary to establish failed exit nodes. These are exit nodes in the penultimate column that have not lead to exit nodes in the right-most column. These exit nodes are easily established as part of the reverse pass. First, any exit node in the penultimate column encountered during the reverse pass is marked as accessible. Then, after the reverse pass is complete, one loop over the penultimate column can be used to check which exit nodes have not been marked as accessible -- these are the failed exit nodes.

Once failed exit nodes have been established, a final reverse pass of the block can be used to prune any successor paths that lead to these failed exit nodes. This is achieved by deleting the successor from the root of any tree which only leads to failed exit nodes.

\subsubsection{Main cause of failure}
\label{sec:other-failures}

\begin{figure}[t]
  \centering
  \includegraphics[width=0.8\columnwidth]{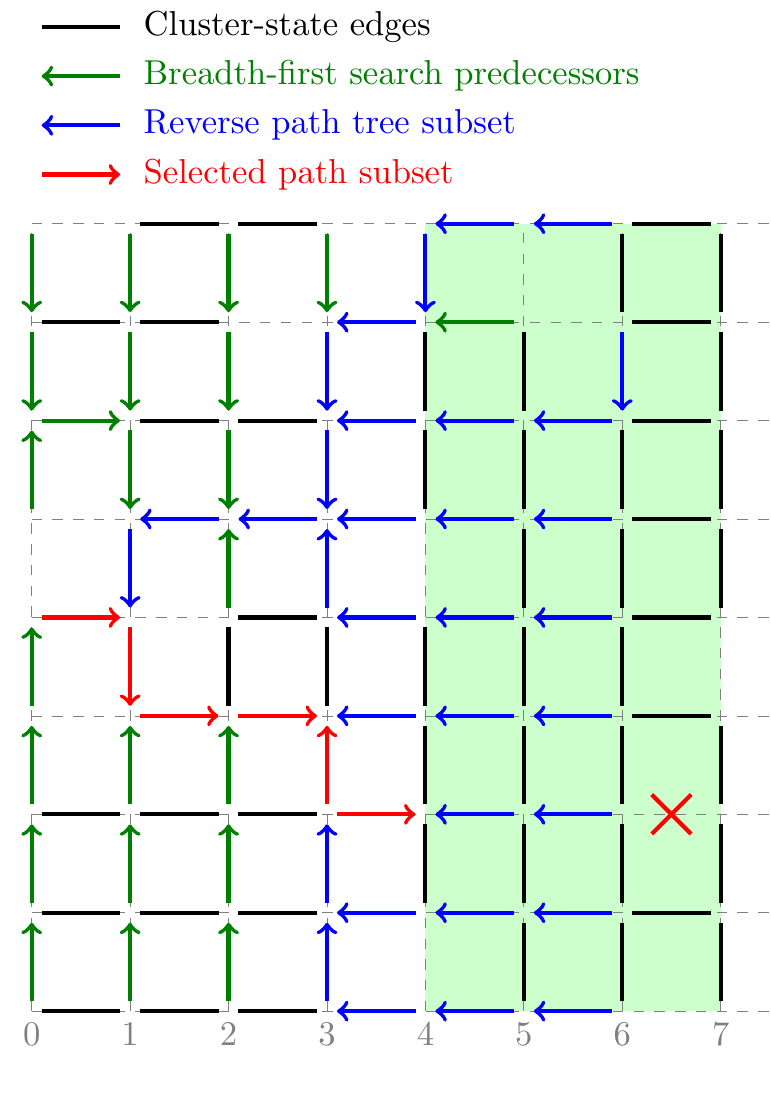}
  \caption[Common failure case in the IBFS]{Diagram showing the most common failure case in the IBFS algorithm. Even though the path can clearly be extended, the algorithm is not able to extend the path because it cannot ``see'' a way around the dead-end, shown as the red cross. This is because the reverse-path tree is missing valid edges that could be used in the path). The problem is due to the inability of IBFS to re-write the path predecessors more than once. Some kind of local search may be required to avoid this error.}
  \label{fig:failure-case}
\end{figure}

There are a number of other issues that arise in the implementation of IBFS. The one that ultimately causes the version of the algorithm presented here to fail is related to the inability to revisit nodes during the BFS phase of the forward pass. This is a direct consequence of not deleting all the search data from the previous search block.

The main failure case is depicted in Figure~\ref{fig:failure-case}, which we found occurs almost immediately for nearly all combinations of search parameters. It happens when IBFS finds a horizontal path through a fully connected region of cluster state. In this case, along this section, all path predecessors point backwards to the left (the blue arrows in Figure~\ref{fig:failure-case}). If a column is reached that is missing a horizontal segment to extend this path, then IBFS will fail, even though the path could extend up or down in order to circumvent the missing horizontal link.

This failure occurs because the BFS algorithm does not ``know'' about edges above and below the horizontal line, that it could use to avoid this missing edge, because it cannot revisit the nodes along the path from different directions. In the GBFS algorithm, these opportunities to extend the path above and below would have been found, because the algorithm is able to rewrite the reverse path tree in the whole block when a new column is added.

This deficiency dramatically reduces the effectiveness of the IBFS algorithm, as we show in the next section. A solution to this problem would require a modification to the BFS process in the IBFS algorithm. Whatever modification is necessary will increase the algorithmic complexity of the solution, but may still represent an improvement of GBFS.

\section{Emulating the search algorithms}
\label{sec:emul-search-algor}

We have written software to emulate the algorithms discussed in the previous sections~\cite{mbqcsim-repo}, for the purpose of deriving timing constraints that may be present in hardware implementations. We used a 2D lattice, as shown in Figure~\ref{fig:block-search}a, with (independently) randomly generated edges following a uniform binary distribution with edge probability $p$. The emulation records the number of memory-related operations that are performed in the course of the algorithm, such as the number of times a distance is written in GBFS.  Memory access tends to be the main bottleneck in many real-time architectures \cite{williams2009roofline}. The results of this emulation are described below. All results were obtained using a graph of height $H=20$. In each graph, each point represents the result of averaging 1000 runs of the emulator for each combination of parameters $p$ (the $x$-axis) and $B$ (the legend).

\begin{figure}[t]
  \centering
  \includegraphics[width=\columnwidth]{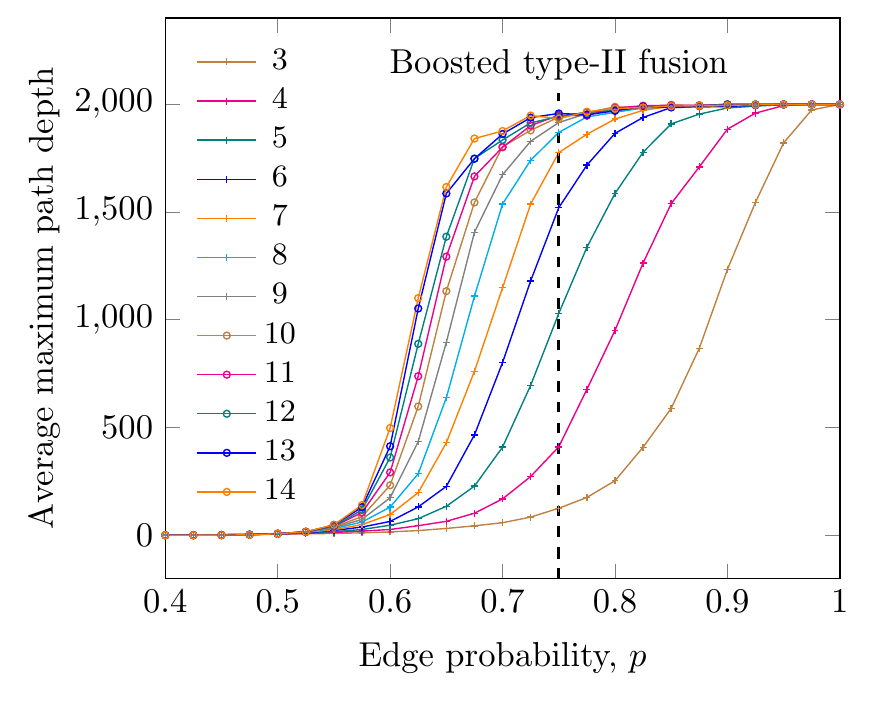}  
  \caption[Average maximum path depth achieved using GBFS]{Average maximum path depth achieved using GBFS, as a function of the block width parameter (given in the legend), for cluster states with varying edge probabilities. The graph shows that the block width has a significant impact on the achievable depth, but there is limited benefit available from arbitrarily increasing the block width. The upper bound depth 2000 is due to only simulating cluster states of width 2000. The vertical dashed line shows the edge probability achieved by using boosted type-II fusion gates to generate the cluster state. This line may be used to establish what block width is necessary to achieve a particular target depth.}
  \label{fig:max-depth-gbfs}
\end{figure}

Figure~\ref{fig:max-depth-gbfs} shows that the GBFS algorithm works, reproducing the results of \cite{morley2017physical}. Specifically, when the edge probability $p > 0.5$, the chance of paths existing increases strongly~\cite{morley2017physical}, and the GBFS algorithm is able to find these paths for block widths $B=5$ to $10$. The graph also shows the success probability of the boosted type-II fusion gate~\cite{gimeno2015three}, which is one possible method to produce the photonic entanglement required for the cluster state. The levelling off of the graphs at depth 2000 is an artefact due to the limit of our simulation. What is important is that, for large enough $B$, the depth approaches 2000 (the limit).

\begin{figure}[t]
  \centering
  \includegraphics[width=\columnwidth]{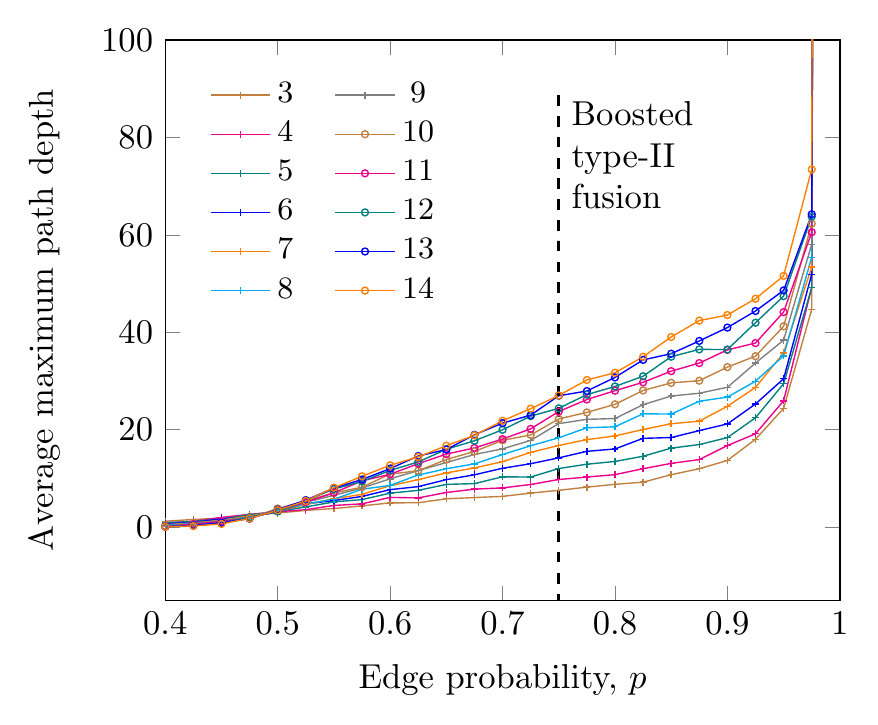}  
  \caption[Average maximum path depth achieved using IBFS]{Average maximum path depth achieved using IBFS, as a function of the block width parameter (given in the legend), for cluster states with varying edge probabilities. The graph shows that the IBFS algorithm performs substantially worse than GBFS for nearly all edge probabilities, due to the limitations outlined in Section~\ref{sec:incr-breadth-first}. This experiment was performed alongside GBFS using the same cluster state width 2000.}
  \label{fig:max-depth-ibfs}
\end{figure}

In contrast, the IBFS algorithm does not work, as shown by Figure~\ref{fig:max-depth-ibfs}. The failure is due to the issue identified in Section~\ref{sec:other-failures}, that viable paths are excluded by the mechanism we used for calculating successors. The failure exists across all edge probabilities $p$ and block widths $B$, apart from the degenerate case $p=1$. It may be possible to modify this algorithm into one that works without too much difficulty; for example, a further (local) search step in the calculation of successors may help the algorithm avoid the primary failure mode.

\begin{figure}[t]
  \centering
  \includegraphics[width=\columnwidth]{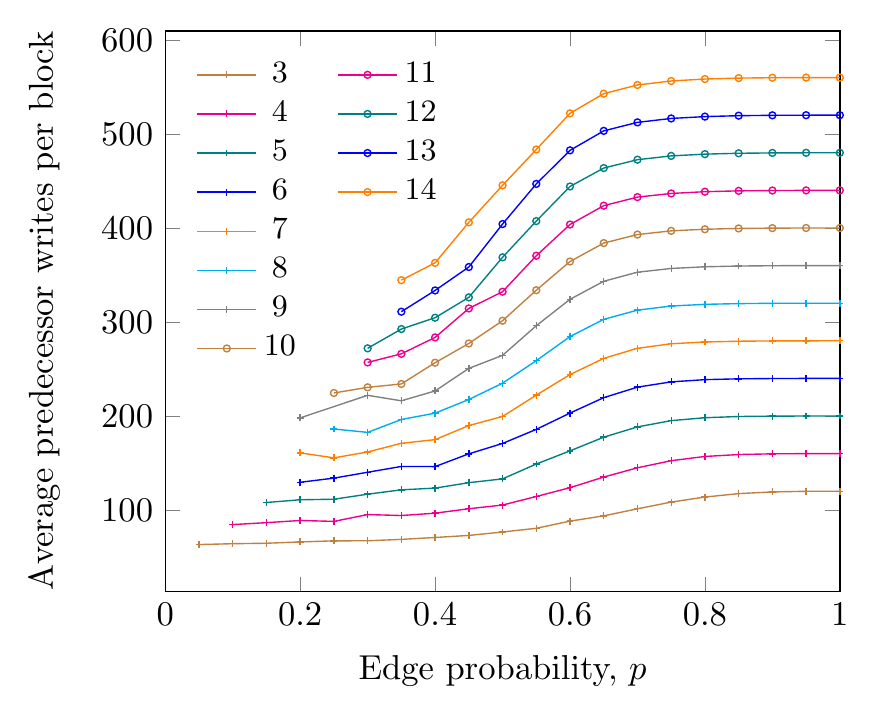}  
  \caption[Average predecessor writes when using GBFS]{Average number of predecessors written to memory during each block search process when using GBFS. At lower edge probabilities, fewer predecessors are written because less of the block may be accessible. As the edge probability increases, the number of predecessors written approaches $2HB$, where $H$ is the cluster height and $B$ is the block width. The factor of two is due to the need to clear the ring buffer at the start of the search process.}
  \label{fig:pred-writes-gbfs}
\end{figure}

The main purpose of the emulation is to estimate the number of computational operations that are required in the implementation of these algorithms, in order to derive timing constraints on the photonic quantum computer. Here, we focus on memory-related operations. Figure~\ref{fig:pred-writes-gbfs} shows the average number of times a predecessor is written into the ring buffer during each photonic clock cycle, in the execution of the GBFS algorithm. As the block width $B$ increases, the number of writes increases, because the graph is larger.

As the edge probability $p$ approaches 1, the number of writes asymptotically approaches a constant $2BH$, because there is a higher chance that the whole graph will be visited twice by the algorithm (once in the resetting step, and a second time in the calculation of the predecessors).

\begin{figure}[t]
  \centering
  \includegraphics[width=\columnwidth]{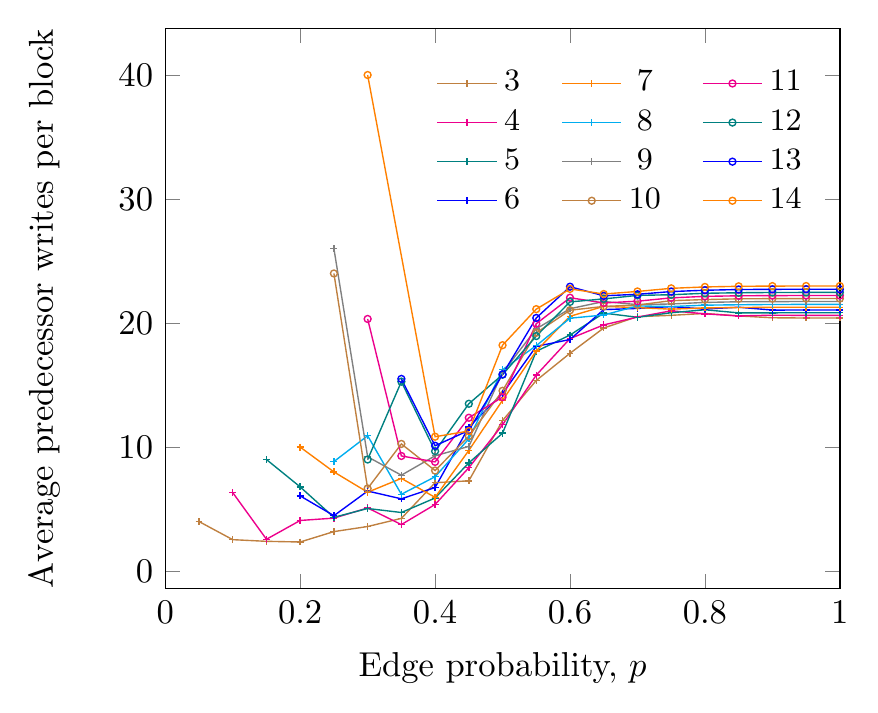}  
  \caption[Average predecessor writes when using IBFS]{Average number of predecessors written to memory during each block search process when using IBFS. Compared to GBFS, the predecessor-write overhead is substantially reduced. At higher edge probabilities, the average number of distance writes is equal to $H$ (the cluster height, fixed at 20 in this experiment), and does not scale with the block width. This is because, on average, only the final column of the block is searched in each IBFS block search process. As the edge probability decreases, the memory overhead increases, because there is a chance that the search process will have to visit previously inaccessible columns in the inner part of the block, that has been made available by the addition of new edges in the final column.}
  \label{fig:pred-writes-ibfs}
\end{figure}

The graph of predecessor writes per block for the IBFS algorithm is shown in Figure~\ref{fig:pred-writes-ibfs}. Compared to GBFS, the IBFS algorithm does not require more predecessor writes for larger block widths, because only the new part of the graph at the right hand side of the green region (Figure~\ref{fig:block-search}a) is searched at each photonic clock cycle. As a result, as the edge probability $p\to 1$, the number of writes per block approaches $H$ (only one column). This improvement compared to GBFS justifies the investigation of these types of algorithms; however, more work is required to find a variant that works properly.

\section{Timing implications for photonic quantum computers}
\label{sec:timing-impl-phot}

The $y$-axis of Figures~\ref{fig:pred-writes-gbfs}~and~\ref{fig:pred-writes-ibfs} in Section~\ref{sec:emul-search-algor} may be interpreted in the context of memory latency for a target memory technology used to implement the ring buffer. In the model discussed here, the memory accesses are performed sequentially, and must all be completed within the photonic clock cycle.
If the photonic clock cycle is $T_p$, and the average number of predecessor writes per block-search is $W_\text{pred}$, then the average maximum acceptable memory write time $t_\text{write}$ is given by
\begin{equation}
t_\text{write} = \frac{T_p}{W_\text{pred}}.
\end{equation}
For example, if the photonic cycle time is \SI{1}{\nano\second}, and the edge probability $p$ is taken as the type-II fusion probability (\SI{75}{\percent}), then a block width of $B=5$ (required to achieve a path depth of approximately 1000, from Figure~\ref{fig:max-depth-gbfs}) would lead to a maximum acceptable write time of $t_\text{write}=\SI{5}{\pico\second}$ (corresponding to 200 predecessor writes). This is an extremely tight timescale in which to achieve a memory write in a digital system. In real-time systems, memory performance is highly dependent on the specific problem being solved and the memory architecture used~\cite{mittal2021survey}; for reference, a recent high-performing device achieved memory latencies on the order~\SI{150}{\pico\second}~\cite{9065587}.

Taking the FPGA in our previous work~\cite{scott2022timing} as an example, memory switching times for distributed RAM are on the order of \SI{0.5}{\nano\second}~\cite{ds182} -- two orders of magnitude too slow for the implementation of the GBFS algorithm discussed here. This means that it would likely not be feasible to implement the ring-buffer-based control system using this FPGA; a higher performance device, or an ASIC, would be required.

For the GBFS algorithm, in the limit of high edge probabilities (the asymptote visible at each block width in Figure~\ref{fig:pred-writes-gbfs}), it is possible to provide a specific formula for the maximum acceptable latency, in terms of implementation parameters of the system:
\begin{equation}
t_\text{write} \approx \frac{T_p}{2BH},
\end{equation}
where $H$ is the cluster-state height and $B$ is the block width.

Performing the same calculation as before, the IBFS algorithm would lead to a maximum acceptable memory write time of $t_\text{write}=\SI{50}{\pico\second}$. Although this is a substantial improvement compared to the $\SI{5}{\pico\second}$ of GBFS, it is still some way off the \SI{150}{\pico\second} latency discussed above. Moreover, if the IBFS algorithm is modified to fix the failure mode discussed previously, it is likely that some additional timing overhead will be introduced in the process (potentially bringing it more in line with the computational cost of GBFS).

\section{Discussion}
\label{sec:discussion}

Throughout the literature on photonic quantum computing based on MBQC, the breadth-first search algorithm and its variants are considered as the primary mechanism for finding paths through cluster states~\cite{Kieling2007,gimeno2015three,morley2017physical}. It is argued that the efficient classical complexity of these algorithms (i.e. polynomial time) makes them suitable for use in the control system for the quantum computer~\cite{Kieling2007,herr2018local}. We show here that this is not the case, due to the large number of memory operations that must be performed (in absolute terms). It is certain that, at the very least, these algorithms must be aggressively optimised and/or parallelised, if they are to be considered viable contenders for the solution of this problem.

However, it is much more likely that entirely different approaches must be adopted for control systems in MBQC-based quantum computing. For example, it may be possible to cast the problem in the framework of in-memory computation, often more appropriate for memory-intensive real-time systems, which may remove some of the overhead inherent in moving data about in the control system~\cite{mittal2021survey}. However, due attention should be paid to fundamental timing constraints, for example arising from trace lengths inside FPGAs, or wires connecting different parts of the digital system, which are often on the order \SI{100}{\pico\second}~\cite{9065587}. Only a few of these delays in a serial system are necessary before a \SI{1}{\giga\hertz} photonic clock rate is not feasible.

Modern approaches to photonic quantum computing do not remove these problems; rather, they modify the specific algorithms involved. For example, in fusion-based quantum computation~\cite{2101.09310}, it is no longer necessary to search for paths through cluster states. Instead, it is necessary to implement an error-correction-like procedure to implement quantum computations in a fault-tolerant framework. In this case, the complexity of these algorithms in a particular implementation model must be analysed, especially regarding the memory overheads involved, to show that timing constraints relating to the overall photonic system~\cite{2103.08612} are met.

\section{Conclusion}
\label{sec:conclusion}

In this work, we have demonstrated that classical control systems impose a significant overhead on photonic implementations of quantum computing, and these overheads must be accounted for in control system designs. In particular, the probabilistic generation of cluster states requires sophisticated real-time graph traversal algorithms to be implemented within a photonic clock cycle. We showed through two different implementations of these path-finding algorithms that there is an inherent tradeoff between implementation speed and path-finding accuracy. Moreover, failure cases (such as shown in Figure~\ref{fig:failure-case}) are inherently tied to the implementation details and need to be accounted for in the design emulation phase. Path-finding, as we show here, is an intrinsically memory-intensive algorithm and the memory access speeds (both read and write) ultimately become the limiting bottleneck on the speed with which computations can be performed in this system.

There are two main implications of this work for photonic quantum computing: the first is that the design of classical control systems needs to be given equal footing in the architectural layout of photonic quantum computers. More precisely, without a concrete control system specification, it is hard to verify the system level constraints that will eventually occur in these systems. However, by picking a concrete control system algorithm (for example, GBFS using $B=5$ and $H = 20$), it follows that a memory technology capable of a read/write time of \SI{150}{\pico\second} likely places a lower bound on the photonic clock period of \SI{30}{\nano\second}, purely due to the digital processing requirement. This is substantially higher than is often considered in the context of photonic quantum computing~\cite{2103.08612,Bourassa2021}.

A second theme is the need to construct explicit implementations of control systems, including algorithms specified in terms familiar to classical computer architects. These explicit proposals for control system algorithms must be specified as a prerequisite step before tackling the much more difficult problem of finding an appropriate digital design, including questions of pipelining, parallelisation, interface latencies, and choice of memory technologies~\cite{hennessy2011computer}. 

The analog-photonic interface between the digital system will impose additional constraints on the quantum computer. Adding the time needed for digital operations to the analog-digital conversion time required for the detectors, and digital-analog conversion time needed for stabilising the analog voltages on the modulators, provides a lower bound on the photonic clock cycle time. Estimating each of these quantities precisely is a critical next step in system level specifications of photonic MBQC, and these methods can be readily extended to any architecture for photonic quantum computing.

\section*{Acknowledgements}

The authors would like to thank Lana Mineh, Naomi Solomons, and Oliver Thomas for reading through the manuscript and providing valuable suggestions. JRS received funding from the Bristol Quantum Engineering Center for Doctoral Training, EPSRC Grant No. EP/L015730/1.
KCB would like to thank the European Research Council for funding support (ERC-StG SBS3-5, 758843).

\appendices

\section{Ring-buffer model for counting memory operations}
\label{sec:ring-buffer-model}

A ring buffer is an implementation of a first-in, first-out (FIFO) data structure~\cite{kruse1998data} which consists of a bounded buffer region whose ends are logically connected, as shown in Figure~\ref{fig:full-system}c. The advantage of this structure is that it may be implemented simply in hardware or software by utilising a contiguous block of memory, and storing the next available location for writing (one past the \texttt{head} in Figure~\ref{fig:full-system}c), and the last valid location for reading (the \texttt{tail}). When data is added to the buffer, the \texttt{head} is incremented once (advances one position anticlockwise in Figure~\ref{fig:full-system}c), and when data is read, \texttt{tail} is incremented once. This way, old data is continually overwritten by new data, and no error occurs provided that \texttt{tail} is always strictly in front of \texttt{head}.

The advantage of block-based MBQC as shown in Figure~\ref{fig:block-search}a is that the block is a fixed size, so the buffer need only be as large as the block width ($B+1$, to account for the possibility of write-before-read). In addition, even though the block subwindow logically moves to the right in Figure~\ref{fig:block-search}a, appearing to require the rewriting of all data at each new block, the ring-buffer model means that each column is only written once. Instead of moving the data, the \texttt{head} and \texttt{tail} pointers are moved, and old column data is overwritten by new column data as the \texttt{head} pointer moves anti-clockwise around the buffer.

Each entry in the buffer shown in Figure~\ref{fig:full-system}c stores a column of block information, and its associated data. This includes the vertical edge data for that column, and the horizontal edge data connecting one column to the next. In addition, the buffer must also store local information required by the implementation of the various algorithms involved in the control system: searching for paths, mapping the measurement pattern, and computing updates to byproduct operators.

In order to simplify the implementation as much as possible, we constrain the classical algorithms to only use data that is compatible with storage in the ring buffer. This means the following two requirements must be satisfied: 
\begin{itemize}
\item The algorithm data must be storable in a way that is distributed across the ring buffer, one data structure per cluster qubit location.
\item The data structure must be the same at all cluster-qubit positions in the ring buffer.
\end{itemize}

Both requirements are intended to simplify a hardware realisation of the control system model as much as possible (e.g. using an FPGA). The first requirement removes the need to consider another data structure in addition to the ring buffer for the storage of algorithm data. The second requirement guarantees straightforward alignment of the ring buffer in memory (by requiring that each buffer location be the same size), which ensures that hardware logic for processing buffer entries does not have to depend on which entry is being read.

In the C++ software~\cite{mbqcsim-repo}, the ring buffer is modelled by the \texttt{NodeWindow} class in \texttt{src/node-window.hpp}. The data stored in the ring buffer has a type constructed using \texttt{MakeNode} (\texttt{src/make-node.hpp}), which amalgamates different classes defined in files of the form \texttt{src/*-node.hpp}. This corresponds to the data shown in Table~\ref{tab:gbfs-data}.

\section{Local pattern rules defining measurement operations}
\label{sec:local-pattern-rules}

The measurement pattern is encoded in a set of local pattern rules, described briefly in Section~\ref{sec:pattern-generation}, which are designed to be stored in the ring buffer. Pattern rules are required to simplify the column measurement implementation as much as possible. In making the measurements, it is necessary to step through each cluster qubit in the red column (Figure~\ref{fig:full-system}a), set its measurement basis (including adaptive measurement setting), make the measurement, and then use the outcome to update byproduct operators. This process is simplified if all these measurements can be made in parallel, and each is fully controlled by information that is local to the cluster qubit being measured (and does not involve, for example, the collection and processing of information stored at multiple nodes). By making each measurement use identical information, the speed of the measurement process in a digital implementation may be maximised, by ensuring that no measurement takes longer than any of the others.

\begin{table}[t]
  \centering
  \begin{tabularx}{\columnwidth}{l lX@{}}
    \toprule
    Data & Meaning\\
    \midrule
    $z$ & Flag indicates $Z$-measurement or $xy$-plane basis \\
    $\theta$ & Base angle (only for $xy$-plane measurement)\\
    $R_s$ & Adaptive measurement setting rule, $(r,s)$\\
    $R_b$ & Byproduct operator update rule, $(r,s)$ \\
    \bottomrule
  \end{tabularx}
  \caption[Data required for local measurement-pattern rules]{Table showing the data required for storing the measurement pattern. The pattern is stored as a set of local rules (one per cluster qubit), which completely specify how each cluster qubit should be measured, and what should be done with the measurement outcome. The pattern rules are generated after a path extension has been found (see Figure~\ref{fig:pattern-gen}).}
  \label{tab:local-pattern-data}
\end{table}

The ring-buffer data required for storing local pattern information is summarised in Table~\ref{tab:local-pattern-data}, and described in the following sections.

\subsubsection{Byproduct operator update rules}
\label{sec:bypr-oper-update}

The measurement pattern rules make reference to the pair $(x,z)$, which is the running value of the byproduct operators $(x_{2k},z_{2k})$. This is updated as the measurement pattern is evaluated by XORing measurement outcomes into either the $x$ or $z$ term in the pair, using rules defined here.

As shown in Figure~\ref{fig:cut-out-pattern}, the measurement outcome $m$ from an on-path qubit $a_n$ is XORed into $x$ or $z$ depending on whether $n$ is even or odd:
\begin{equation}
  (x,z) \mapsto \begin{cases}
    (x,z\oplus m) & \text{if $n$ is even},\\
    (x\oplus m, z) & \text{if $n$ is odd}.
\end{cases}
\end{equation}
This rule is stored as a pair $R_b = (r,s)$ which is either $(1,0)$ or $(0,1)$, depending on whether the outcome should be added to $x$ or $z$ respectively. The value of this pair for a qubit $a_n$ is denoted $R_b(a_n)$.

For each on-path qubit $a_n$ whose measurement outcome $m$ is added to a term in $(x,z)$, the measurement outcome from any adjacent cut-out qubit must also be added to that same term (see Figure~\ref{fig:cut-out-pattern} for the special case where the $a_n$ lie along a horizontal line). This leads to the rule that the outcome $m$ from a cut-out qubit $b$ may be added to either of the terms $(x,z)$ multiple times, because $b$ may be adjacent to multiple on-path qubits. To account for this, each cut-out qubit stores a pair $R_b = (r,s)$, which expresses the net effect of this cut-out qubit on the byproduct operators, when the measurement outcome from this cut-out is $m$:
\begin{equation}
  (x,z) \mapsto (x\oplus mr,z\oplus ms).
\end{equation}
This pair is obtained for a particular cut-out qubit $b$ by adding (pairwise modulo-2) all the values $R_b(a)$ for on-path qubits $a$ ($a\in P$) that are adjacent to $b$ ($a\sim b$):
\begin{equation}
  \label{eq:cut-out-combinations}
  R_b(b) = \bigoplus_{\substack{a\sim b\\a\in P}} R_b(a).
\end{equation}
This calculation is performed by the inner-most loop in Figure~\ref{fig:pattern-gen}.

\subsubsection{Measurement basis angle and dependency rules}
\label{sec:adapt-meas-sett-1}

\begin{figure*}[t]
  \centering
  \includegraphics[width=0.8\textwidth]{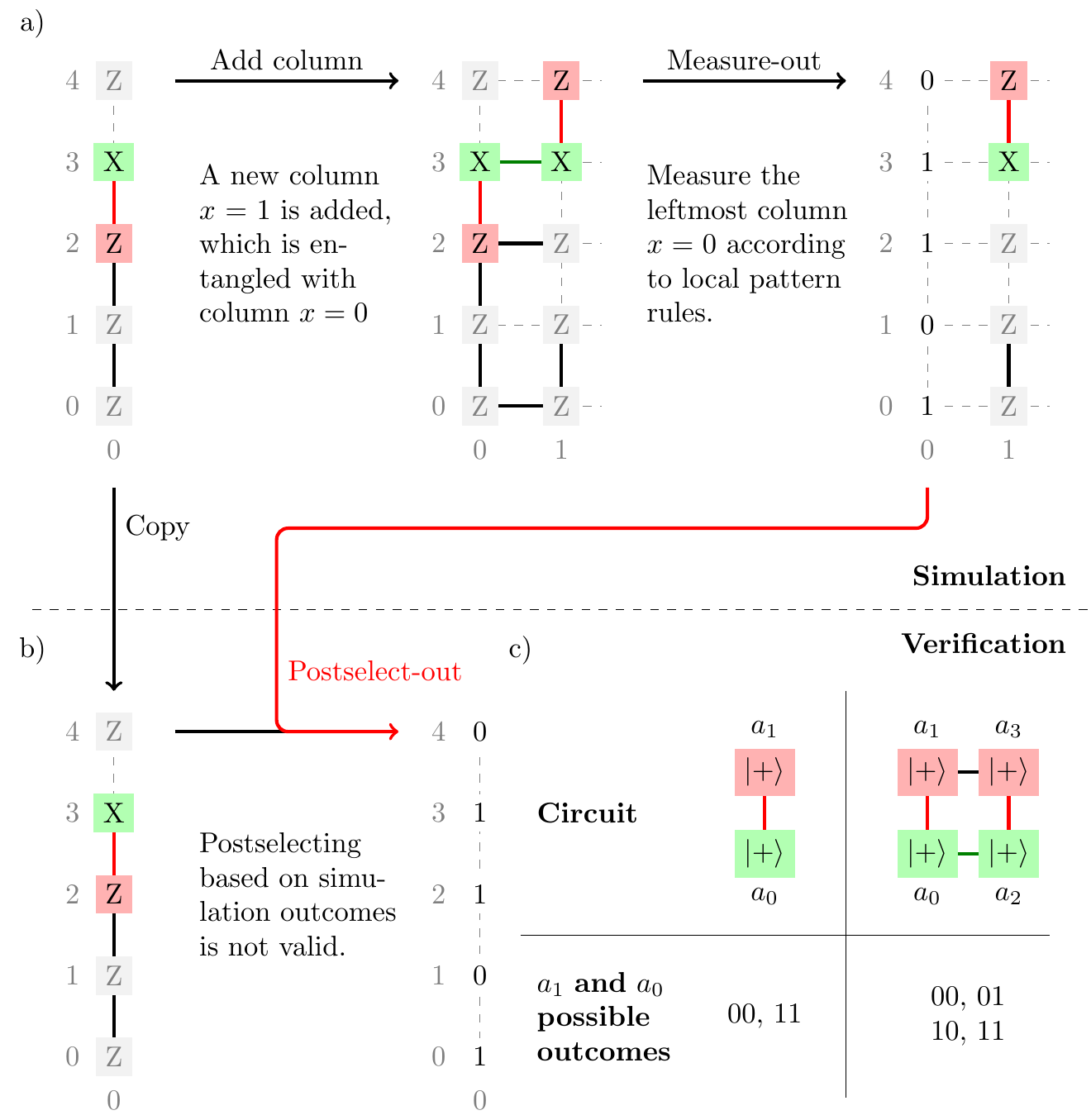}
  \caption[Cluster-state simulation and verification]{a) Review of the main simulation procedure, where a column is added to the right, and then the leftmost column is measured-out according to local pattern rules. b) The ``ideal'' method of verification, where outcomes are made to agree between the simulation and verification by postselecting the verification column. c) A minimal example showing how the incompatibility arises in a simple case. In both circuits shown, $a_1$ and $a_0$ are measured in the $Z$- and $X$-bases, and the possible outcomes are shown below the circuits.}  
  \label{fig:sim-verify}
\end{figure*}

Each qubit in the cluster state is either measured in the $Z$-basis (if it is a cut-out qubit, or if it is not directly connected to the measurement pattern), or the $xy$-plane, for all other measurements. A flag $z$ is stored in each cluster qubit node to specify in which basis it is measured.

The measurement basis angle for each $xy$-plane measurement is stored as a base angle $\theta$, and an adaptive measurement setting $s$. The base angle is shown inside the green filled boxes in Figure~\ref{fig:cut-out-pattern}, and the measurement setting is shown as the input in the bottom-left corner of each square. For example, the second on-path qubit in Figure~\ref{fig:cut-out-pattern} has $\theta=-\phi_1$, and $s$ is the current value of the byproduct operator $z_{2k}$ just before the measurement of that qubit. The base angle is a static property of the pattern (it does not depend on any measurement outcomes or byproduct operators), and relates to what $R_x$ and $R_z$ rotations are realised by the pattern.

The adaptive measurement setting rule $R_s(a_n)$ is stored as a pair $(r,s)$, which describes how to calculate the adaptive measurement setting $s$ from the current values of the byproduct operators $(x,z)$:
\begin{equation}
  s = R_s(a_n)\cdot (x,z) = rx \oplus sz.
\end{equation}

In the C++ software, the local pattern rules described above are stored in the ring buffer, in data defined by the \texttt{PatternNode} class (see \texttt{src/pattern-node.hpp}). One copy of this data structure is stored at each location in the ring buffer, each corresponding to a single cluster qubit.

\section{Verification of the validity of the model}
\label{sec:simul-volt-noise}

The emulated implementation of the model quantum computing system presented in this paper was verified to ensure that it is correct. This was achieved by performing a quantum simulation of the entire system, including all the algorithms presented in this paper, and comparing the fidelity of the output state of the simulated logical qubit (in the MBQC measurement pattern) to the state of a reference qubit that has undergone the same logical one-qubit operation.

Figure~\ref{fig:sim-verify}a shows a summary of the steps involved in the simulation. It is only necessary to store two columns of entangled cluster qubits, independent of the block width or the width of the cluster state\footnote{This follows from a similar argument to the one used in \cite[II.D]{Raussendorf2003} for concatenating measurement patterns, which shows that the measurements may be interleaved with the entangling operations.}. In each ``simulation round'', a new column is added on the right and entangled with the column on the left. Then, the column on the left is measured and removed according to the local pattern rules constructed using the algorithm in Figure~\ref{fig:pattern-gen} (see also Appendix~\ref{sec:local-pattern-rules}). On a laptop with \SI{8}{\gibi\byte} memory, this permits the simulation of systems up to $H=14$ (although $H=10$ is a more practical upper limit from the point of view of execution time). There is no limit on the total number of columns in the simulation, nor on the block width $B$, apart from the simulation time which is proportional to the total cluster state width. The simulator~\cite{mbqcsim-repo} is based on the quantum simulator QSL~\cite{qsl-repo,lana-thesis}, which contains a specialist resizing simulator for the purpose of efficiently performing the operations shown in Figure~\ref{fig:sim-verify}a~\cite[Chapter 4]{john-thesis}.

The output of the simulation (after all columns have been simulated) can be compared to the reference logical qubit. However, it is desirable to verify the simulation at all columns of the cluster state, so as to be able to locate where an error occurred (in what might be a very long cluster state).

In the ideal verification scheme, the measurement pattern would be truncated just after the column being verified (i.e. before the entanglement step), and would be postselected to have the same measurement outcomes as the main simulation, so as to leave one qubit that would act as the ``final'' qubit of the measurement pattern. This proposal is shown in the copy/postselect process between Figure~\ref{fig:sim-verify}a and Figure~\ref{fig:sim-verify}b.

However, it is not necessarily possible to postselect the verification column (Figure~\ref{fig:sim-verify}b) using the measurement outcomes from the main simulation, for the reason shown by the minimal example in Figure~\ref{fig:sim-verify}c. In that case, the verification column $a_1a_0$ is the state $\ket{\psi}=\ket{0{+}} + \ket{1{-}}$ (normalisation is omitted). If qubit $a_1$ is measured in the $Z$-basis, and $a_0$ is measured in the $X$-basis, then the only possible outcomes are 00 or 11. However, when $a_2$ and $a_3$ are entangled as shown (all the lines in the figure are $CZ$ gates), the state on all four qubits $a_3a_2a_1a_0$ becomes
\begin{equation}
  \ket{\psi} = \ket{{+}00{+}} + \ket{{-}01{-}} + \ket{{-}10{-}} + \ket{{+}11{+}}.
\end{equation}
When $a_0$ and $a_1$ are measured in the same bases as before, all four outcomes 00, 01, 10, and 11 are possible. This latter set of possibilities reflects a larger set of potential outcomes from the main simulation. Therefore, it may not be possible to enforce the same measurement outcomes in the verification column.

\begin{figure}[t]
  \centering
  \includegraphics[width=\columnwidth]{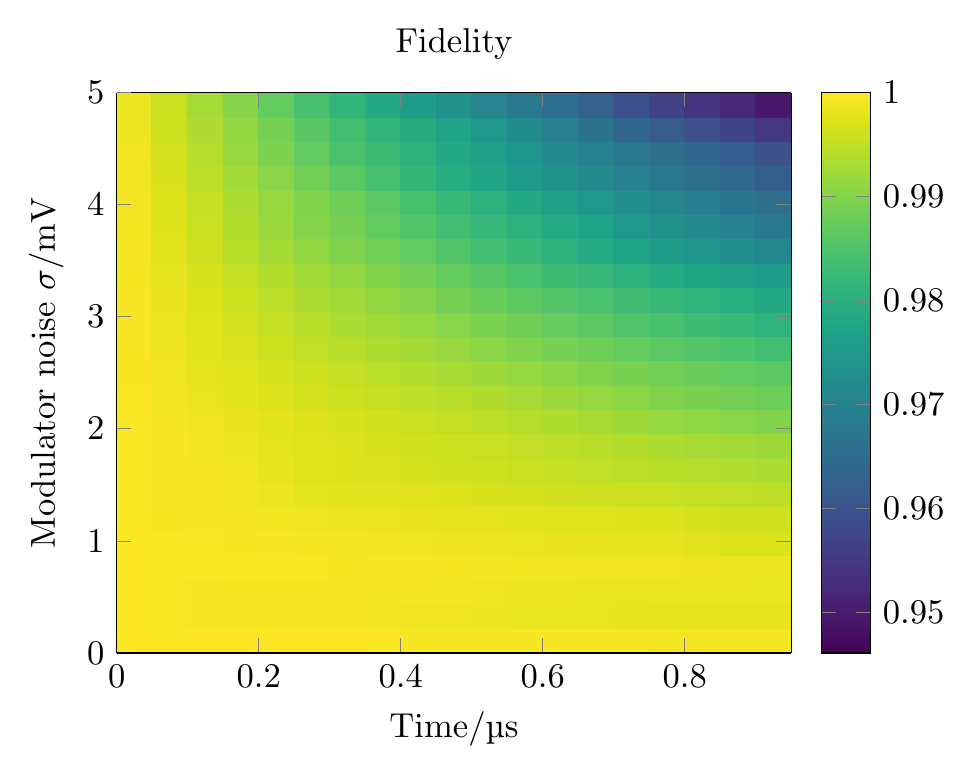}
  \caption[Average fidelity as a function of analog voltage noise and time]{The heatmap shows the average fidelity of the logical qubit realised using the identity path pattern, as a function of elapsed time (derived from a photonic cycle time of \SI{1}{\nano\second}), and the standard deviation of the Gaussian white noise in the modulator voltage. The experiment was conducted by simulating a pathfinding process using GBFS in cluster state of height $H=7$, for a range of noise levels. Each experiment was repeated 1000 times and the results averaged. The simulation assumes $V_\pi=\SI{1}{\volt}$ for the modulators that set the measurement bases.}
  \label{fig:error-with-time}
\end{figure}

As a result, it is necessary to allow the verification columns (the copied columns shown in Figure~\ref{fig:sim-verify}b) to be measured without constraints. This may lead to outcomes that differ from the main simulation, in this particular measurement round. However, since the main simulation and verification measurement outcomes only diverge in the current simulation round, the verification does check that the columns strictly before the current column are correct. It therefore serves equally well as a column-by-column verification of the entire pattern.

The ``output'' from each verification column is a single unmeasured qubit, which represents the end of the one-qubit path. The state of this qubit is compared with a reference qubit that has undergone the same logical operations that the quantum computing system is supposed to have performed. If the two states agree ($\text{fidelity}=1$), then the emulator is verified. 

A side effect of column-by-column verification is the ability to introduce noise into the measurement bases shown in Figure~\ref{fig:full-system}b, and investigate the effect of these errors on the fidelity of the output state. An example of this analysis is shown in Figure~\ref{fig:error-with-time}, for the identity gate. This line of inquiry may be extended to a numerical (computational) analysis of errors in MBQC systems, which may complement more theoretical analysis of error correction schemes.

The cluster state simulation is performed by the \texttt{ClusterSim} class in \texttt{src/cluster-sim.hpp}~\cite{mbqcsim-repo}. Each simulation round is performed by the \texttt{simulate} member function of the \texttt{PathSim} class (\texttt{src/path-sim.hpp}), which also performs the verification steps.

\section{Reproducing the results}
\label{sec:reproducing-results}

The results in the paper may be reproduced on a Linux-based computer by obtaining the mbqcsim git repository~\cite{mbqcsim-repo} (\texttt{master} branch, commit \texttt{cc5a0bf5}), compiling the main C++ library according to the instructions in the \texttt{README}, and then running the script \texttt{scripts/paper-figures.py}. Figures~\ref{fig:max-depth-gbfs}, \ref{fig:max-depth-ibfs}, \ref{fig:pred-writes-gbfs}, \ref{fig:pred-writes-ibfs} and \ref{fig:error-with-time} contain seed-based random elements. The seeds used for each of the figures in this paper are contained in the file \texttt{python/mbqcsim/paper.py}, which also shows what parameter combinations were used to reproduce the results.

The failure case shown in Figure~\ref{fig:failure-case} may be reproduced by running the following command, which shows a step-by-step breakdown of the evolving pathfinding algorithm:
{\ttfamily\small
\begin{verbatim}
$ pathf -s3996592937216137949 -d -aibfs \
  -p0.9 -B4 -H9 -W10000
\end{verbatim}
}
The \texttt{pathf} program is built as part of the compilation process in the git repository. For more information about the programs \texttt{pathf} and \texttt{esim}, see the \texttt{README}.

\bibliographystyle{apsrev4-1-mod}
\bibliography{ref}

\end{document}